\documentclass[showpacs,preprintnumbers,amsmath,amssymb,twocolumn]{revtex4}

\usepackage{graphicx}
\usepackage{dcolumn}
\usepackage{bm}

\begin{document}

\title{Superfluidity in a Three-flavor Fermi Gas with SU(3) Symmetry}
\author{\normalsize{Lianyi He, Meng Jin and Pengfei Zhuang}}
\affiliation{Physics Department, Tsinghua University, Beijing
100084, China}

\begin{abstract}
We investigate the superfluidity and the associated
Nambu-Goldstone modes in a three-flavor atomic Fermi gas with
SU(3) global symmetry. The s-wave pairing occurs in flavor
anti-triplet channel due to the Pauli principle, and the
superfluid state contains both gapped and gapless fermionic
excitations. Corresponding to the spontaneous breaking of the
SU(3) symmetry to a SU(2) symmetry with five broken generators,
there are only three Nambu-Goldstone modes, one is with linear
dispersion law and two are with quadratic dispersion law. The
other two expected Nambu-Goldstone modes become massive with a
mass gap of the order of the fermion energy gap in a wide coupling
range. The abnormal number of Nambu-Goldstone modes, the quadratic
dispersion law and the mass gap have significant effect on the low
temperature thermodynamics of the matter.
\end{abstract}


\maketitle

\section {Introduction}
\label{s1}
The superfluidity in strongly interacting atomic Fermi gas and the
associated BCS-Bose Einstein condensation(BEC) crossover
phenomena\cite{BEC1,BEC2,BEC3} have been observed in
experiments\cite{exp1,exp2,exp3,exp4} via the method of Feshbach
resonance. The experimental study of superfluidity in atomic Fermi
gas may be important for us to understand the solid-state
phenomena such as high-temperature superconductivity, and may give
some clue to search for the ground state of the dense quark matter
and nuclear matter. In the past years, most theoretical and
experimental studies concentrated on the two-flavor systems such
as a $^6$Li gas with the two lowest hyperfine states (In this
paper, we use the word ``flavor'' in particle physics to denote
the internal degrees of freedom of the fermionic atoms). Compared
to electrons in solids, atomic systems offer more internal degrees
of freedom. For alkali atoms, nuclear spin $I$ and electron spin
$S$ are combined in a hyperfine state with total angular momentum
$F$. While typical electronic systems are constrained to a SU(2)
spin rotational symmetry, the total angular momentum $F$ can be
larger than $1/2$, resulting in $2F+1$ hyperfine states differing
by their azimuthal quantum number $m_F$. Therefore, the atomic
Fermi gas can provide us a way to study the superfluidity with
broken symmetry higher than the $U(1)$ one. In this paper, we will
focus on a three-flavor system with a SU(3) global symmetry. Such
a system has been investigated in some
works\cite{SU3-1,SU3-2,SU3-3}.

It is well-known that, associated with the spontaneous breaking of
a global symmetry, there should be corresponding
Nambu-Goldstone(NG) bosons. Such NG-bosons dominate the low
temperature thermodynamics of the system. According to the
Goldstone theorem\cite{NG1,NG2}, if an internal continuous
symmetry group is spontaneously broken down to a subgroup with $N$
broken generators, $N$ NG-bosons appear in Lorentz-invariant
systems, i.e., the number of NG-bosons is equal to the number of
broken generators. However, from the Nielsen-Chadha(NC)
theorem\cite{NC}, for systems without Lorentz invariance the
number of NG-bosons can be less than the number of broken
generators. Let $N_1$ and $N_2$ be the numbers of gapless
excitations which have, respectively, the dispersion laws
$\omega\sim|\vec{p}|$ and $\omega\sim|\vec{p}|^2$ in the limit of
long wavelength, the number of broken generators satisfies the
relation $N<N_1+2N_2$. For the equality between the number of
NG-bosons and the number of the broken generators, there is an
important criterion: If $\langle[Q_i,Q_j]\rangle=0$ for any two
broken generators $Q_i$ and $Q_j$, $i,j=1,2,...,N$, the number of
NG-bosons is equal to the number of the broken
generators\cite{Tomas}.

For the three-flavor Fermi gas with SU(3) symmetry we will
consider in this paper, the ground state of the system contains
both gapped and gapless fermionic excitations. When the SU(3)
symmetry is spontaneously broken to a SU(2) subgroup with five
broken generators, we will show with an explicit calculation that
there are only three NG-modes. Among them, one has linear
dispersion law and the other two have quadratic dispersion law.
The reason for the abnormal number of NG-modes and the appearance
of quadratic dispersion law is found to be the fact that, the
condition $\langle[Q_i,Q_j]\rangle=0$ is not satisfied due to the
density imbalance between the gapped and gapless fermions.

The abnormal number of NG-modes and the non-linear dispersion law
have been widely discussed in relativistic field theory at finite
density\cite{Tomas,AG1,AG2}. They were also found in the study of
two flavor color superconductivity in the Nambu--Jona-Lasinio
model\cite{NJL1} where the condition $\langle[Q_i,Q_j]\rangle=0$
is not satisfied due to the lack of color neutrality. However, the
abnormal number of NG-bosons can not be realized in superfluid
quark matter and has no observable effect, since the color
neutrality should be imposed via some mechanism such as gluon
condensation and the NG-bosons should be eaten up by the gluons
via the Higgs mechanism. In atomic Fermi gas, there is no
constraint like the color neutrality, and the NG-modes are
physical degrees of freedom which dominate the low temperature
thermodynamics of the system. The theoretic prediction of the
NG-modes may be tested in future experiments via the measurement
of the thermodynamic quantities. In addition, the mass gap of the
two massive collective modes found in \cite{NJL1} is very small
compared with the quark energy gap, while the corresponding mass
gap in the three-flavor Fermi gas is of the order of the fermion
energy gap, which makes remarkable effect on the low temperature
thermodynamics.

The paper is organized as follows. In Section \ref{s2}, we set up
the model for the three-flavor Fermi gas with SU(3) global
symmetry. In Section \ref{s3}, we investigate the ground state of
the system and the symmetry breaking. In Section \ref{s4}, we
investigate the pair fluctuation around the superfluid ground
state, identify the NG-modes and find their dispersion laws. In
Section \ref{s5}, we discuss whether we can recover the five
NG-modes. We summarize in Section \ref{s6}. The natural unit of
$c=\hbar=k_B=1$ is used through the paper.

\section {The Model}
\label{s2}
The physical system we are interested in in this paper is an idea
system composed of three flavors of fermions with attractive
interaction. Such a system can be realized in cold atomic Fermi
gas such as a $^6$Li or $^{40}$K gas where the three flavors come
from three degenerate hyperfine states\cite{three}. Generally, the
system can be modelled by the Lagrangian density
\begin{equation}
{\cal
L}=\psi^\dagger\left(i\partial_t+\frac{\nabla^2}{2m}+\mu\right)\psi+{\cal
L}_{int},
\end{equation}
where $\psi\equiv(\psi_1,\psi_2,\psi_3)^T$ and
$\psi^\dagger\equiv(\psi_1^*,\psi_2^*,\psi_3^*)$ are the
three-component fermion fields, $\mu$ is their chemical potential,
and $m$ is their mass. We have assumed that the chemical
potentials of three flavors are the same due to the chemical
equilibrium.

It is generally believed that the s-wave channel is the dominant
pairing channel. According to the Pauli principle, the total wave
function of the Cooper pair must be anti-symmetric. Using the
decomposition $3\otimes3=\bar{3}\oplus6$ for the SU(3) group, the
s-wave pairing must be associated with the anti-triplet channel in
flavor space. In this paper we consider only the s wave pairing
channel, the interaction can be modelled by
\begin{equation}
{\cal L}_{int}=\frac{g}{4}(\psi^*_\alpha i\epsilon_{\alpha\beta
I}\psi^*_\beta)(\psi_{\alpha^\prime}i\epsilon_{\alpha^\prime\beta^\prime
I}\psi_{\beta^\prime}),
\end{equation}
where $g$ is the bare coupling related to the s-wave scattering
length and $\epsilon_{ijk}$ is the total anti-symmetric tensor.
Throughout, summation is implicit over repeated flavor index. Note
that the interaction Lagrangian can also be written as
\begin{equation}
{\cal
L}_{int}=\frac{g}{4}\sum_{a=2,5,7}(\psi^\dagger\lambda_a\psi^*)(\psi^T\lambda_a\psi),
\end{equation}
where $\lambda_a(a=1,2,...,8)$ are the Gell-mann matrices. The
model Lagrangian has the symmetry SU(3)$\otimes$ U(1), i.e., it is
invariant under the transformation
\begin{equation}
\psi\rightarrow e^{-iT_a \theta_a}\psi, \ \ a=0,1,2,...,8,
\end{equation}
where $T_0=I_3$ is the generator of the $U(1)$ group and
$T_a=\frac{\lambda_a}{2}(a=1,2,...,8)$ are the generators of the
SU(3) group. Due to the above symmetry, the system possesses nine
conserved charges or generators $Q_a(a=0,1,2,...,8)$ given by
\begin{equation}
Q_a=\int d^3{\bf x}\psi^\dagger T_a \psi.
\end{equation}

Like the two-flavor or $U(1)$ system, for attractive coupling $g$
we can perform an exact Stratonovich-Hubbard transformation to
introduce the pair fields
\begin{equation}
\Phi_I\sim \frac{g}{2}(\psi_\alpha i\epsilon_{\alpha\beta
I}\psi_\beta),\ \ \Phi_I^*\sim \frac{g}{2}(\psi^*_\alpha
i\epsilon_{\alpha\beta I}\psi^*_\beta)
\end{equation}
for $I=1,2,3$. With the Nambu-Gorkov fields defined as
\begin{equation}
\Psi=\left(\begin{array}{c} \psi
\\ \psi^* \end{array}\right)\ ,\ \ \ \Psi^\dagger=\left(\begin{array}{cc}
\psi^\dagger& \psi^T\end{array}\right)\ ,
\end{equation}
the partition function $Z$ of the system can be expressed as
\begin{equation}
Z=\int[d\Psi^\dagger][d\Psi][d\Phi^*_I][d\Phi_I]e^{\int_x
\left(\frac{1}{2}\Psi^\dagger{\cal K}\Psi-{\Phi^*_I\Phi_I\over
g}\right)}
\end{equation}
with the kernel ${\cal K}[\Phi_I^*,\Phi_I]$ defined as
\begin{equation}
{\cal K}[\Phi^*_I,\Phi_I]=\left(\begin{array}{cc}
-\partial_\tau+\frac{\nabla^2}{2m}+\mu&i\epsilon_{\alpha\beta
I}\Phi_I
\\ i\epsilon_{\alpha\beta I}\Phi^*_I&-\partial_\tau-\frac{\nabla^2}{2m}-\mu\end{array}\right)
\end{equation}
in the imaginary time ($\tau=it$) formalism of finite temperature
field theory with $\int_x=\int_0^\beta d\tau\int d^3{\bf x}$,
where $\beta$ is the inverse of temperature, $\beta=1/T$.
Integrating out the fermionic degrees of freedom, we obtain
\begin{equation}
Z=\int[d\Phi^*_I][d\Phi_I]e^{-S_{eff}[\Phi^*_I,\Phi_I]}
\end{equation}
with the effective action
\begin{equation}
S_{eff}[\Phi^*_I,\Phi_I]=\int_x\frac{\Phi^*_I\Phi_I}{g}-\frac{1}{2}\text
{Tr}\ln{\cal K}[\Phi^*_I,\Phi_I].
\end{equation}

\section {The Ground State}
\label{s3}
At some critical temperature $T_c$ the system should undergo the
phase transition from the normal phase with the SU(3) symmetry to
the superfluid phase where the SU(3) symmetry is spontaneously
broken. Since we focus on the low temperature region where $T\ll
T_c$, the mean field approximation or saddle point approximation
is believed to be a good treatment for the ground state. The order
parameters which characterize the superfluid phase or symmetry
broken phase are defined as the expectation values of the pair
fields
\begin{equation}
\Delta_I=\langle\Phi_I\rangle,\ \ \
\Delta_I^*=\langle\Phi_I^*\rangle,\ \ I=1,2,3.
\end{equation}

Let us consider the homogeneous and isotropic superfluid state
where the order parameters are independent of the coordinates. The
thermodynamic potential $\Omega$ in mean field approximation can
be expressed as
\begin{eqnarray}
\Omega&=&\frac{1}{\beta
V}S_{eff}[\Phi^*_I=\Delta^*_I,\Phi_I=\Delta_I]\\
&=&\frac{\Delta^*_I\Delta_I}{g}-\frac{1}{2\beta}\sum_n\int
\frac{d^3{\bf p}}{(2\pi)^3}\text{Tr} \ln {\cal
G}^{-1}(i\omega_n,{\bf p}),\nonumber
\end{eqnarray}
where $V$ is the volume of the system and ${\cal G}^{-1}$ is the
inverse of the fermion propagator in momentum space:
\begin{equation}
{\cal G}^{-1}(i\omega_n,{\bf p})=\left(\begin{array}{cc}
i\omega_n-\xi_p&i\epsilon_{\alpha\beta I}\Delta_I
\\ i\epsilon_{\alpha\beta I}\Delta^*_I&i\omega_n+\xi_p\end{array}\right)
\end{equation}
with $\omega_n$ the Matsubara frequency for fermions and
$\xi_p=p^2/(2m)-\mu$. A straightforward algebra shows that the
determinate of ${\cal G}^{-1}$ in the Nambu-Gorkov$\otimes$flavor
space reads
\begin{equation}
\det{\cal
G}^{-1}=\left[(i\omega_n)^2-\xi_p^2\right]\left[(i\omega_n)^2-\xi_p^2-\Delta^2\right]^2
\end{equation}
which indicates that the thermodynamic potential depends only on
the quantity $\Delta^2$ defined as
\begin{equation}
\Delta^2=|\Delta_1|^2+|\Delta_2|^2+|\Delta_3|^2.
\end{equation}
The fermionic excitation spectra can be determined by $\det{\cal
G}^{-1}=0$. For positive chemical potential, the superfluid phase
contains both gapped and gapless fermionic excitations,
\begin{equation}
\omega_{1,2}({\bf p})=\pm\sqrt{\xi_p^2+\Delta^2},\ \ \omega_3({\bf
p})=\xi_p.
\end{equation}
With the quasiparticle dispersions, the thermodynamic potential
can be evaluated as
\begin{eqnarray}
\Omega&=&-\frac{m\Delta^2}{4\pi a_s}-\Delta^2\int\frac{d^3{\bf
p}}{(2\pi)^3}\left(\frac{1}{E_p+\xi_p}-\frac{1}{2\epsilon_p}\right)\\
&-&\frac{1}{\beta}\int\frac{d^3{\bf
p}}{(2\pi)^3}\left[2\ln(1+e^{\beta E_p})+\ln(1+e^{\beta
\xi_p})\right],\nonumber
\end{eqnarray}
where we have defined the notation $E_p=\sqrt{\xi_p^2+\Delta^2}$
and replaced the bare coupling $g$ by the low energy limit of the
two body T-matrix
\begin{equation}
\frac{m}{4\pi a_s}=-\frac{1}{g}+\int\frac{d^3{\bf
p}}{(2\pi)^3}\frac{1}{2\epsilon_p}
\end{equation}
with $a_s$ the s-wave scattering length and $\epsilon_p=p^2/(2m)$.

Now we discuss the symmetry breaking pattern. Similar to the two
flavor color superconductivity, the symmetry breaking pattern in
the current case is
\begin{equation}
SU(3)\otimes U(1)\longrightarrow SU(2)\otimes \tilde{U}(1).
\end{equation}
To see the broken and unbroken symmetry groups explicitly, it is
convenient for us to choose
\begin{equation}
\label{choice}\Delta_1=\Delta_2=0, \ \ \ \Delta_3\equiv\Delta\neq
0
\end{equation}
without loss of generality. In this case, only flavors $1$ and $2$
participate in the Cooper pairing and flavor $3$ remains unpaired,
a SU(2) subgroup with generators $T_1,T_2,T_3$ and a
$\tilde{U}(1)$ subgroup with generator
$\tilde{T}_0=(\sqrt{3}T_0-T_8)/2$ remain unbroken, and the broken
generators are $T_4,T_5,T_6,T_7$ and
$\tilde{T}_8=(T_0+\sqrt{3}T_8)/2$. We should emphasis that all the
physical results do not depend on the specific choice of symmetry
breaking direction due to the fact that the Lagrangian is
invariant under the SU(3) transformation.

To determine physical quantities in the superfluid state, we
should solve the gap equation together with the number equation.
Assuming the total number density $n$ is fixed, we can introduce
the Fermi momentum $p_F$ and Fermi energy $\epsilon_F$ through the
definitions $n=3\times p_F^3/(6\pi^2)$ and
$\epsilon_F=p_F^2/(2m)$. At mean field level, the gap equation
which determines the energy gap $\Delta$ can be derived via
$\partial \Omega/\partial \Delta=0$, namely,
\begin{equation}
-\frac{m\Delta}{4\pi a_s}=\Delta\int\frac{d^3{\bf
p}}{(2\pi)^3}\left[\frac{1-2f(E_p)}{2E_p}-\frac{1}{2\epsilon_p}\right],
\end{equation}
and the number equation can be derived via $n=-\partial
\Omega/\partial \mu$, namely,
\begin{equation}
n=\int\frac{d^3{\bf
p}}{(2\pi)^3}\left[\left(1-\frac{\xi_p}{E_p}\right)+2\frac{\xi_p}{E_p}f(E_p)+f(\xi_p)\right],
\end{equation}
where $f(x)$ is the Fermi-Dirac distribution function.  For the
specific choice of symmetry breaking direction (\ref{choice}), the
number densities $n_1,n_2$ for the paired flavors and $n_3$ for
the unpaired flavor can be evaluated as
\begin{eqnarray}
&& n_1=n_2=\int\frac{d^3{\bf
p}}{(2\pi)^3}\left[\frac{1}{2}\left(1-\frac{\xi_p}{E_p}\right)+\frac{\xi_p}{E_p}f(E_p)\right],\nonumber\\
&& n_3=\int\frac{d^3{\bf p}}{(2\pi)^3}f(\xi_p),
\end{eqnarray}
which satisfy $n=n_1+n_2+n_3$. Once the Cooper pairing occurs, the
number of the paired fermions becomes different from the number of
the unpaired fermions, $n_1=n_2\neq n_3$. This difference can be
parameterized by the ratio $\alpha$ defined as
\begin{equation}
\alpha=\frac{n_1+n_2-2n_3}{n}.
\end{equation}

\begin{figure}[!htb]
\begin{center}
\includegraphics[width=7cm]{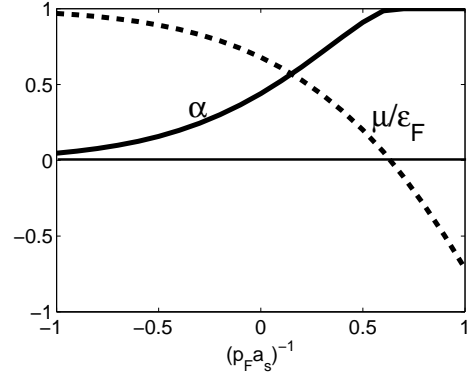}
\caption{The chemical potential $\mu$ scaled by the Fermi energy
$\epsilon_F$ and the ratio $\alpha$ as functions of
$(p_Fa_s)^{-1}$ at $T=0$. \label{fig1}}
\end{center}
\end{figure}

Solving the coupled set of gap equation and number equation, we
can obtain the gap $\Delta$, the chemical potential $\mu$ and the
ratio $\alpha$ as functions of the coupling $(p_Fa_s)^{-1}$.
Before the detailed numerical calculation, we give a qualitative
estimate at $T=0$. In the BCS limit, the energy gap $\Delta$ is
very small and the chemical potential is approximately the Fermi
energy, the ratio $\alpha$ will be very small. In the BEC limit,
the chemical potential becomes negative and the unpaired fermions
disappear, the ratio approaches to the limit $\alpha=1$. In
Fig.\ref{fig1}, we show the numerical results of the chemical
potential $\mu$ and the ratio $\alpha$ at $T=0$. The ratio
$\alpha$ is always positive which means $n_1=n_2>n_3$.

What does the nonzero ratio $\alpha$ mean? To answer this
question, we calculate the expectation values of the generators
$Q_a(a=1,2,...,8)$. They can be calculated via the formula
\begin{equation}
\langle Q_a\rangle=\frac{V}{2\beta}\sum_n\int \frac{d^3{\bf
p}}{(2\pi)^3}\text {Tr}\Big[\tau_3\otimes T_a{\cal
G}(i\omega_n,{\bf p})\Big],
\end{equation}
where $\tau_3$ is the third Pauli matrix in the Nambu-Gorkov
space. The explicit form of the fermion propagator in the
Nambu-Gorkov$\otimes$flavor space can be evaluated as
\begin{equation}
{\cal G}(i\omega_n,{\bf p})=\left(\begin{array}{cccccc} {\cal
G}_-^\Delta&0&0&0&-i{\cal F}&0
\\ 0&{\cal G}_-^\Delta&0&i{\cal F}&0&0\\ 0&0&{\cal G}_-^0&0&0&0 \\ 0&-i{\cal F}&0&{\cal G}_+^\Delta&0&0 \\ i{\cal F}&0&0&0&{\cal G}_+^\Delta&0 \\
0&0&0&0&0&{\cal G}_+^0
\end{array}\right)
\end{equation}
with the nonzero matrix elements defined as
\begin{eqnarray}
&&{\cal
G}_\pm^\Delta=\frac{i\omega_n\mp\xi_p}{(i\omega_n)^2-E_p^2},\ \ \
\ \ {\cal
G}_\pm^0=\frac{1}{i\omega_n\pm\xi_p},\nonumber\\
&&{\cal F}=\frac{\Delta}{(i\omega_n)^2-E_p^2}.
\end{eqnarray}
After a straightforward matrix algebra, we find
\begin{eqnarray}
&&\langle Q_a\rangle=0,\ \ a=1,2,...,7,\nonumber\\
&&\frac{\langle
Q_8\rangle}{V}=\frac{n_1+n_2-2n_3}{\sqrt{3}}=\frac{\alpha
n}{\sqrt{3}}.
\end{eqnarray}
According to the commutation relation of SU(3) group, we have
\begin{eqnarray}
\langle[Q_4,Q_5]\rangle=\langle[Q_6,Q_7]\rangle=i\sqrt{3}\langle
Q_8\rangle=i\alpha nV\neq0.
\end{eqnarray}
Therefore, the nonzero ratio $\alpha$ in the superfluid phase
means that, the condition $\langle[Q_i,Q_j]\rangle=0$ which is
sufficient for the equality between the number of NG-bosons and
the number of the broken generators is not satisfied in such a
system. However, we now can not conclude that the number of
NG-bosons is not equal to the number of broken generators.

\section {The Nambu-Goldstone Modes}
\label{s4}
We investigate now the pair fluctuations around the superfluid
state and examine whether there are five NG-modes. If there exist
five NG-modes, they must be the collective modes associated with
the pair fluctuations. After the field shift $\Phi_3\rightarrow
\Phi_3+\Delta$, the effective action reads
\begin{equation}
S_{eff}=\int_x\frac{\Phi^*_I\Phi_I}{g}-\frac{1}{2}\text
{Tr}\ln\left[{\cal G}^{-1}+\Sigma(\Phi^*_I,\Phi_I)\right]
\end{equation}
with the matrix $\Sigma$ defined as
\begin{eqnarray}
\Sigma=\left(\begin{array}{cccccc} 0&0&0&0&i\Phi_3&-i\Phi_2
\\ 0&0&0&-i\Phi_3&0&i\Phi_1\\ 0&0&0&i\Phi_2&-i\Phi_1&0 \\ 0&i\Phi_3^*&-i\Phi_2^*&0&0&0 \\ -i\Phi_3^*&0&i\Phi_1^*&0&0&0 \\
i\Phi_2^*&-i\Phi_1^*&0&0&0&0
\end{array}\right).
\end{eqnarray}

To identity the existence of NG-modes and study their dispersion
laws, we need to evaluate the effective action only to the
quadratic terms of the pair fields. Using the derivative
expansion, the effective action up to the quadratic terms can be
expressed as
\begin{equation}
S_{eff}=\int_x\frac{\Phi^*_I\Phi_I}{g}+\frac{1}{4}\text
{Tr}\left[{\cal G}\Sigma(\Phi^*_I,\Phi_I){\cal
G}\Sigma(\Phi^*_I,\Phi_I)\right].
\end{equation}

To do calculations in momentum space, we define the Fourier
transformations
\begin{eqnarray}
\Phi_I(x)&=&\frac{1}{\sqrt{\beta V}}\sum_q
e^{-i\nu_n\tau+i{\bf q}\cdot{\bf x}}\Phi_I(q),\nonumber\\
\Phi_I^*(x)&=&\frac{1}{\sqrt{\beta V}}\sum_q e^{-i\nu_n\tau+i{\bf
q}\cdot{\bf x}}\Phi^*_I(q)
\end{eqnarray}
for the pair fields, where the four momentum in the imaginary time
formalism is defined as $q=(i\nu_n,{\bf q})$ with $\nu_n$ the
Matsubara frequency for bosons. We have
$\Phi^*_I(-q)=(\Phi_I(q))^*$ due to the above definitions. After a
straightforward matrix algebra, the effective action of the pair
fields can be decomposed into three parts:
\begin{equation}
S_{eff}[\Phi_I^*,\Phi_I]=S_1[\Phi_1^*,\Phi_1]+S_2[\Phi_2,\Phi_2]+S_3[\Phi_3,\Phi_3].
\end{equation}
The pair fields for $I=1,2,3$ do not mix with each other, which
makes the calculation of NG-modes quite easy. Each part of the
effective action $S_I$ takes the form
\begin{eqnarray}
S_I&=&\frac{1}{2}\sum_q\Big[\Pi_I^{11}(q)\Phi_I^*(-q)\Phi_I(q)+\Pi_I^{22}(q)\Phi_I^*(q)\Phi_I(-q)\nonumber\\
&+&\Pi_I^{12}(q)\Phi_I^*(-q)\Phi_I^*(q)+\Pi_I^{21}(q)\Phi_I(q)\Phi_I(-q)\Big],
\end{eqnarray}
where the functions $\Pi_I^{ij}(q)$ for $I=1,2$ take the same form
due to the residue SU(2) symmetry,
\begin{eqnarray}
\Pi_I^{11}(q)&=&\Pi_I^{22}(-q)=\frac{1}{g}\nonumber\\
&+&\frac{1}{2\beta V}\sum_p\left[{\cal G}_+^0(p){\cal
G}_-^\Delta(p+q)+{\cal
G}_+^\Delta(p){\cal G}_-^0(p+q)\right],\nonumber\\
\Pi_I^{12}(q)&=&\Pi_I^{21}(q)=0,
\end{eqnarray}
and the functions $\Pi_3^{ij}(q)$ are given by
\begin{eqnarray}
\Pi_3^{11}(q)&=&\Pi_3^{22}(-q)=\frac{1}{g}+\frac{1}{\beta
V}\sum_p\left[{\cal
G}_+^\Delta(p){\cal G}_-^\Delta(p+q)\right],\nonumber\\
\Pi_3^{12}(q)&=&\Pi_3^{21}(q)=\frac{1}{\beta V}\sum_p\left[{\cal
F}(p){\cal F}(p+q)\right].
\end{eqnarray}
The functions $\Pi_I^{ij}$ for $I=1,2,3$ are evaluated in
Appendixes A and B. To identify the existence of the NG-modes and
determine their dispersion laws, it is convenient to use the real
and imaginary parts of the complex pair fields defined as
\begin{eqnarray}
&&\Phi_I(x)=\left(\varphi_I(x)+i\phi_I(x)\right)/\sqrt{2},\nonumber\\
&&\Phi_I^*(x)=\left(\varphi_I(x)-i\phi_I(x)\right)/\sqrt{2}\end{eqnarray}
for $I=1,2,3$. The dispersion laws are determined by the zeros of
the determinate of the matrix $\Pi_I^{ij}$,
\begin{equation}
\Pi_I^{11}(q)\Pi_I^{22}(q)-\Pi_I^{12}(q)\Pi_I^{21}(q)=0.
\end{equation}

\subsection {The $I=1,2$ or $T_4,T_5,T_6,T_7$ Sector}
For $I=1,2$, after the analytical continuation $i\nu_n\rightarrow
q_0+i\varepsilon$, the function $\Pi_I^{11}$ can be expressed as
\begin{eqnarray}
&&\Pi_I^{11}(q_0,{\bf q})=q_0H(q_0,{\bf q})+J(q_0,{\bf q}),
\end{eqnarray}
where the functions $H$ and $J$ are even function of ${\bf q}$ and
hence depend only on ${\bf q}^2$, see Appendix A. Firstly, let us
examine whether there are four gapless NG-modes corresponding to
the broken generators $T_4,T_5,T_6,T_7$. To this end, we take
${\bf q}^2=0$ to calculate the mass gaps of the collective modes.
The mass gaps of the collective modes are given by the roots of
the equation
\begin{eqnarray}
q_0^2H(q_0,{\bf 0})H(-q_0,{\bf 0})=0,
\end{eqnarray}
where we have used the fact $J(q_0,{\bf 0})=0$. Obviously,
$q_0^2=0$ is a root which gives two gapless NG-modes. To examine
whether there exist other two gapless modes, we need to check
whether the equation
\begin{eqnarray}
H(0,{\bf 0})=0
\end{eqnarray}
is satisfied in the superfluid phase. It is easy to find the
interesting relation between $H(0,{\bf 0})$ and $\langle
Q_8\rangle$,
\begin{eqnarray}
H(0,{\bf 0})=-\frac{n_1+n_2-2n_3}{\Delta^2}=-\frac{\sqrt{3}\langle
Q_8\rangle}{\Delta^2V}=-\frac{\alpha n}{\Delta^2}.
\end{eqnarray}
Since $\alpha$ can not be zero once BCS pairing occurs, as we have
shown in the last section, we conclude that, there are only two
gapless NG-modes, and the other two expected NG-modes become
massive.

We now calculate the dispersion law of the gapless NG-modes and
the mass gap of the massive modes. In the low energy limit
$q_0\rightarrow 0,{\bf q}\rightarrow 0$, we can expand the
functions $H$ and $J$ as Taylor series of $(q_0,{\bf q})$ at the
point $(0,{\bf 0})$ and keep only the leading terms. To find the
dispersion law of gapless NG-modes, we take the expansion
\begin{eqnarray}
q_0H(0,{\bf 0})+\frac{{\bf q}^2}{2}\frac{\partial^2J(q_0,{\bf
q})}{\partial {\bf q}^2}\bigg|_{(0,{\bf 0})}=0.
\end{eqnarray}
Due to the relation obtained in Appendix A,
\begin{eqnarray}
\frac{\partial^2J(q_0,{\bf q})}{\partial {\bf q}^2}\bigg|_{(0,{\bf
0})}&=&-\frac{1}{m}H(0,{\bf 0}),
\end{eqnarray}
the NG-modes have a quadratic dispersion law near ${\bf q}^2=0$,
\begin{eqnarray}
q_0=\frac{{\bf q}^2}{2m}.
\end{eqnarray}
It is very interesting that here the quantity $m$ is just the
fermion mass. For the massive modes, we try to find the zero of
the function $H(q_0,{\bf q})$. In the low energy limit, we take
the expansion
\begin{eqnarray}
H(0,{\bf 0})+q_0\frac{\partial H(q_0,{\bf q})}{\partial
q_0}\bigg|_{(0,{\bf 0})}+\frac{{\bf
q}^2}{2}\frac{\partial^2H(q_0,{\bf q})}{\partial
{\bf q}^2}\bigg|_{(0,{\bf 0})}=0,\nonumber\\
\end{eqnarray}
which leads to the dispersion law
\begin{eqnarray}
q_0=m_1+\frac{{\bf q}^2}{2m_2},
\end{eqnarray}
where the mass gap $m_1$ is given by
\begin{eqnarray}
m_1=-H(0,{\bf 0})\left\{\frac{\partial H(q_0,{\bf q})}{\partial
q_0}\bigg|_{(0,{\bf 0})}\right\}^{-1},
\end{eqnarray}
and the quantity $m_2$ is defined as
\begin{eqnarray}
m_2=-\frac{\partial H(q_0,{\bf q})}{\partial q_0}\bigg|_{(0,{\bf
0})}\left\{\frac{\partial^2H(q_0,{\bf q})}{\partial {\bf
q}^2}\bigg|_{(0,{\bf 0})}\right\}^{-1}.
\end{eqnarray}

\begin{figure}[!htb]
\begin{center}
\includegraphics[width=7cm]{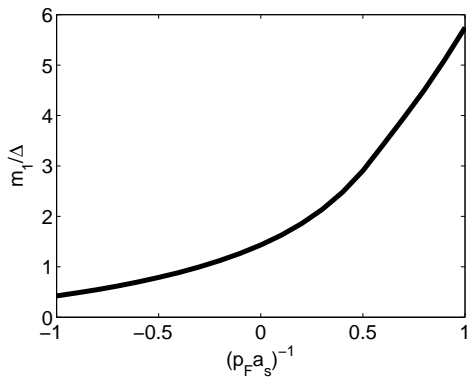}
\caption{The ratio between the mass gap $m_1$ of the massive
collective modes and the energy gap $\Delta$ as a function of
$(p_Fa_s)^{-1}$ at $T=0$. \label{fig2}}
\end{center}
\end{figure}

In the above calculations, we have employed the trick of Taylor
expansion which is valid in the low energy limit. In
Fig.\ref{fig2} we showed the ratio between the mass gap $m_1$ of
the massive collective modes and the energy gap $\Delta$ of the
fermionic excitations as a function of the coupling
$(p_Fa_s)^{-1}$ at $T=0$. In the weak coupling limit, the ratio
approaches to zero, which reflects the fact that the effect of
nonzero $\alpha$ and hence nonzero $H(0,{\bf 0})$ can be
neglected. In this case, we can approximately find five NG-modes
with linear dispersion law in the energy and momentum region
\begin{equation}
m_1<q_0\ll\Delta,\ \ \ m_1<v_F|{\bf q}|\ll\Delta,
\end{equation}
where $v_F=p_F/m$ is the Fermi velocity, and our result is
consistent with the numerical calculation in \cite{SU3-3} where
the authors worked in the weak coupling limit and found five NG
modes. However, in a wide range of the coupling such as
$-1<(p_Fa_s)^{-1}<1$ shown in Fig.\ref{fig2}, the mass gap $m_1$
is of the order of $\Delta$ and hence the order of $T_c$, which
means that the effect of nonzero $\alpha$ can not be neglected and
the mass gap $m_1$ becomes important for the low temperature
thermodynamics.

We conclude that the abnormal number of NG-modes and the mass gap
$m_1$ make sense in a wide range of the coupling and have
significant effect on the low temperature thermodynamics. This
situation is quite different from the case of two flavor color
superconductivity in the Nambu--Jona-Lasinio model where the mass
gap of the massive modes is very small compared with the energy
gap of the quarks\cite{NJL1}.

\subsection {The $I=3$ or $\tilde{T}_8$ Sector}
For $I=3$, the situation is quite conventional. One can check that
the functions take the same form as the ones in the two-flavor or
$U(1)$ system. Let us first identify that there exists a gapless
NG-mode corresponding to the broken generator $\tilde{T}_8$. To
complete this task we need only the explicit form of the functions
$\Pi_3^{ij}$ at $q_0={\bf q}=0$ and check the relation
\begin{equation}
\Pi_3^{11}(0,{\bf 0})\Pi_3^{22}(0,{\bf 0})-\Pi_3^{12}(0,{\bf
0})\Pi_3^{21}(0,{\bf 0})=0.
\end{equation}
Using the explicit form of the functions $\Pi_3^{ij}$ and the gap
equation for $\Delta$, we obtain the relation
\begin{eqnarray}
\Pi_3^{11}(0,{\bf 0})=\Pi_3^{22}(0,{\bf 0})=-\Pi_3^{12}(0,{\bf
0})=-\Pi_3^{21}(0,{\bf 0})
\end{eqnarray}
at any temperature below $T_c$. Hence we have proven that there
must be a gapless NG-mode corresponding to the broken generator
$\tilde{T}_8$.

Next we determine the dispersion law of this NG-mode. Similarly,
we expand the functions $\Pi_3^{ij}(q_0,{\bf q})$ as Taylor series
of $(q_0,{\bf q})$ at the point $(0,{\bf 0})$ and keep only the
leading terms. From the explicit form of the functions
$\Pi_3^{ij}$ evaluated in Appendix B, the Taylor expansions to the
lowest order take the following form
\begin{eqnarray}
&&\Pi_3^{11}(q_0,{\bf q})=A+q_0B+{\bf q}^2D/2,\nonumber\\
&&\Pi_3^{22}(q_0,{\bf q})=A-q_0B+{\bf q}^2D/2,\nonumber\\
&&\Pi_3^{12}(q_0,{\bf q})=-A+q_0^2C/2+{\bf q}^2E/2,\nonumber\\
&&\Pi_3^{21}(q_0,{\bf q})=-A+q_0^2C/2+{\bf q}^2E/2,
\end{eqnarray}
where the coefficients $A,B,C,D,E$ are not explicitly shown. The
dispersion law of the NG-mode is determined by the equation
\begin{eqnarray}
&&\left(A+q_0B+{\bf q}^2D/2\right)\left(A-q_0B+{\bf q}^2D/2\right)\nonumber\\
-&&\left(-A+q_0^2C/2+{\bf q}^2E/2\right)^2=0.
\end{eqnarray}
Keeping the lowest order in $q_0$ and ${\bf q}$, we obtain a
linear dispersion law
\begin{eqnarray}
q_0=v_s|{\bf q}|.
\end{eqnarray}
The velocity of the NG-mode is given by
\begin{eqnarray}
v_s=\sqrt{\frac{A(D+E)}{B^2-AC}}.
\end{eqnarray}
Since the functions $\Pi_3^{ij}$ take the same form as the ones in
the two-flavor system with broken $U(1)$ symmetry, the behavior of
this NG-mode will be the same as the one in the two-flavor
system\cite{jan,liu}.

\section {Can We Recover Five NG Modes?}
\label{s5}
We have shown that in a three-flavor Fermi gas with SU(3) gauge
symmetry, there are only three gapless NG-modes in the superfluid
state. A natural question we may ask is the possibility to recover
the five NG-modes. In this section we will argue that there is no
way to obtain five NG-modes in such a system.

Firstly, one may criticize that the abnormal number of NG-modes
may be due to the specific choice of the symmetry breaking
direction $\Delta_1=\Delta_2=0, \Delta_3\equiv\Delta$. If we take
the following symmetry breaking direction
\begin{eqnarray}
\Delta_1=\Delta_2=\Delta_3=\Delta/\sqrt{3},
\end{eqnarray}
we have automatically $n_1=n_2=n_3$ and hence $\langle
Q_8\rangle=0$, and we may expect five NG-modes with this choice.
However, as we have emphasized, the physical quantities such as
the ground state and the number and dispersion laws of the
NG-modes do not depend on the specific choice of the symmetry
breaking direction. In the symmetric case with
$\Delta_1=\Delta_2=\Delta_3$, the broken and unbroken generators
will be changed, and correspondingly $\langle Q_1\rangle, \langle
Q_4\rangle$ and $\langle Q_6\rangle$ are nonzero. If one works
with this choice, he will certainly get the same number and
dispersion laws of the NG-modes as we have obtained.

Secondly, we may relax the constraint of equal chemical potentials
for the three flavors. For instance, with the choice
$\Delta_1=\Delta_2=0$ and $\Delta_3\equiv\Delta$, we can set
$\mu_1=\mu_2=\mu$ for flavors $1$ and $2$ and $\mu_3=\mu+\mu_0$
for flavor $3$. By requiring $n_1=n_2=n_3$ we can guarantee that
all $\langle Q_a\rangle$ are zero and the condition
$\langle[Q_i,Q_j]\rangle=0$ is satisfied for any $i$ and $j$.
However, in this case, the chemical potentials must be adjusted, a
proper value of $\mu_0\neq 0$ is needed, and the SU(3) symmetry of
the model Lagrangian is explicitly broken with the broken
generators $T_4,T_5,T_6,T_7$. An explicit calculation like the one
in \cite{NJL2} shows that the four NG-modes corresponding to the
$I=1,2$ sector obtain a mass gap $\mu_0$. In the BCS region where
the quantity $\alpha$ in the equal chemical potential system is
small, the corresponding $\mu_0$ is also small. Such a phenomenon
can be regarded as the spontaneous breaking of the approximate
symmetry, and the corresponding collective modes with a small mass
gap $\mu_0$ can be considered as pesudo NG-modes like the case of
color superconductivity in the Nambu--Jona-Lasinio
model\cite{NJL2}. However, since the nonzero $\mu_0$ explicitly
breaks the SU(3) symmetry, a specific choice of the symmetry
breaking direction may be dangerous \cite{NJL3}. In fact, if we
calculate the susceptibilities
\begin{equation}
\frac{\partial^2 \Omega}{\partial
\Delta_1^2}\Big|_{\Delta_1=\Delta_2=0}= \frac{\partial^2
\Omega}{\partial\Delta_2^2}\Big|_{\Delta_1=\Delta_2=0},
\end{equation}
we find they are negative for $\mu_0\neq 0$. This indicates that
if the numbers of the three flavor are fixed, the specific choice
of the pairing pattern is forbidden. In this case, we should solve
the chemical potentials $\mu_1,\mu_2,\mu_3$ and the condensates
$\Delta_1,\Delta_2,\Delta_3$ from a large set of equations.

In conclusion, we can never find five NG-modes in the SU(3)
system. Only in the BCS limit, we can obtain five approximate
NG-modes with linear dispersions via neglecting the effect of
nonzero $\alpha$.

\section {Summary}
\label{s6}
We have investigated the superfluidity and the associated NG-modes
in an atomic Fermi gas with three degenerate hyperfine states. In
our model, the pairing occurs in the s-wave and flavor
anti-triplet channel, and the chemical potentials are constrained
to be equal due to chemical equilibrium. In the superfluid state,
there are both gapped and gapless fermionic excitations, i.e.,
paired and unpaired fermions. Only in the BEC region where the
chemical potential becomes negative, the unpaired fermions
disappear at zero temperature. Once the pairs are condensed, the
SU(3) symmetry is spontaneously broken down to a SU(2) subgroup
with five broken generators. We showed that there are only three
NG-modes, the one corresponding to the diagonal generator is
conventional and has linear dispersion law, and the other two have
quadratic dispersion law. The additional two expected NG-modes
obtain a mass gap. While the mass gap is very small compared with
the energy gap $\Delta$ of the fermions in the BCS limit, it is of
the order of $\Delta$ in a wide range of the coupling
$(p_Fa_s)^{-1}$ and can be reached in experiments of atomic Fermi
gas. As a consequence, the abnormal number of the NG-modes, the
quadratic dispersion law and the mass gap have significant effect
on the low temperature thermodynamics of the three-flavor Fermi
gas.

{\bf Acknowledgments:}\  The work was supported in part by the
grants NSFC10425810, 10435080, 10575058 and SRFDP20040003103.

\begin{widetext}
\appendix
\section {The functions $\Pi_I^{ij}(q)$ for $I=1,2$ }
\label{app1}
In this Appendix we evaluate the functions $\Pi_I^{ij}(q)$ for
$I=1,2$. From the relations $\Pi_I^{22}(q)=\Pi_I^{11}(-q)$ and
$\Pi_I^{12}(q)=\Pi_I^{21}(q)=0$, we need to evaluate $\Pi_I^{11}$
only. After the Matsubara frequency summation and the analytical
continuation $i\nu_n\rightarrow q_0+i\varepsilon$ we obtain
\begin{equation}
\Pi_I^{11}(q)=\frac{1}{g}+\frac{1}{2}\int\frac{d^3{\bf
p}}{(2\pi)^3}\left[\frac{1-f(E_{p})-f(\xi_{p-q})}{q_0-E_{p}-\xi_{p-q}}u_{p}^2+\frac{f(E_{p})
-f(\xi_{p-q})}{q_0+E_{p}-\xi_{p-q}}v_{p}^2+\frac{1-f(E_{p})-f(\xi_{p+q})}{q_0-E_{p}-\xi_{p+q}}u_{p}^2
+\frac{f(E_{p})-f(\xi_{p+q})}{q_0+E_{p}-\xi_{p+q}}v_{p}^2\right]
\end{equation}
with the coherent coefficients $u_p^2$ and $v_p^2$ defined as
$u_p^2=\left(1+\xi_p/E_p\right)/2$ and
$v_p^2=\left(1-\xi_p/E_p\right)/2$. In the superfluid phase with
$\Delta\neq0$, using the gap equation for $\Delta$, we can express
it as
\begin{eqnarray}
&&\Pi_I^{11}(q)=q_0H(q)+J(q)
\end{eqnarray}
with the functions $H(q)$ and $J(q)$ defined as
\begin{eqnarray}
H(q)&=&\int\frac{d^3{\bf
p}}{(2\pi)^3}\frac{1}{2E_{p}}\left[\frac{1-f(E_{p})-f(\xi_{p-q})}{q_0-E_{p}-\xi_{p-q}}
-\frac{f(E_{p})-f(\xi_{p-q})}{q_0+E_{p}-\xi_{p-q}}+\frac{1-f(E_{p})-f(\xi_{p+q})}{q_0-E_{p}-\xi_{p+q}}
-\frac{f(E_{p})-f(\xi_{p+q})}{q_0+E_{p}-\xi_{p+q}}\right],\nonumber\\
J(q)&=&\int\frac{d^3{\bf
p}}{(2\pi)^3}\Bigg[\frac{\xi_p-\xi_{p-q}}{2E_{p}}\left(\frac{1-f(E_{p})-f(\xi_{p-q})}{q_0-E_{p}-\xi_{p-q}}
-\frac{f(E_{p})-f(\xi_{p-q})}{q_0+E_{p}-\xi_{p-q}}\right)\nonumber\\
&&\ \ \ \ \ \ \ \ \ \ \
+\frac{\xi_p-\xi_{p+q}}{2E_{p}}\left(\frac{1-f(E_{p})-f(\xi_{p+q})}{q_0-E_{p}-\xi_{p+q}}
-\frac{f(E_{p})-f(\xi_{p+q})}{q_0+E_{p}-\xi_{p+q}}\right)\Bigg].
\end{eqnarray}

We now list some properties of the functions $H(q)$ and $J(q)$
which are useful to determine the dispersion laws. It is easy to
observe that the functions $H$ and $J$ are both even functions of
${\bf q}$,
\begin{equation}
H(q_0,{\bf q})=H(q_0,-{\bf q}),\ \ \ J(q_0,{\bf q})=J(q_0,-{\bf
q}).
\end{equation}
The function $H(0,{\bf 0})$ can be expressed as
\begin{equation}
H(0,{\bf 0})=-\int\frac{d^3{\bf
p}}{(2\pi)^3}\frac{1}{E_{p}}\left[\frac{1-f(E_{p})-f(\xi_{p})}{E_{p}+\xi_{p}}+\frac{f(E_{p})-f(\xi_{p})}{E_{p}-\xi_{p}}\right].
\end{equation}
By using the identities
\begin{equation}
\frac{1}{E_{p}}\frac{1}{E_{p}+\xi_{p}}=\frac{2v_p^2}{\Delta^2},\ \
\ \frac{1}{E_{p}}\frac{1}{E_{p}-\xi_{p}}=\frac{2u_p^2}{\Delta^2},
\end{equation}
we derive the relation between $H(0,{\bf 0})$ and $\langle
Q_8\rangle$,
\begin{equation}
H(0,{\bf 0})=-\frac{n_1+n_2-2n_3}{\Delta^2}=-\frac{\sqrt{3}\langle
Q_8\rangle}{\Delta^2V}.
\end{equation}
The derivative of the function $H$ with respect to $q_0$ at
$q_0=0,{\bf q}={\bf 0}$ can be written as
\begin{eqnarray}
\frac{\partial H(q_0,{\bf q})}{\partial q_0}\bigg|_{(0,{\bf
0})}&=&-\int\frac{d^3{\bf
p}}{(2\pi)^3}\frac{1}{E_{p}}\left[\frac{1-f(E_{p})-f(\xi_{p})}{(E_{p}+\xi_{p})^2}-\frac{f(E_{p})-f(\xi_{p})}{(E_{p}-\xi_{p})^2}\right].
\end{eqnarray}
The function $J(q)$ can be expressed as
\begin{eqnarray}
J(q)&=&-\frac{{\bf q}^2}{2m}H(q)+\frac{1}{m}\int\frac{d^3{\bf
p}}{(2\pi)^3}\frac{{\bf p}\cdot{\bf
q}}{2E_{p}}\Bigg[\frac{1-f(E_{p})-f(\xi_{p-q})}{q_0-E_{p}-\xi_{p-q}}
-\frac{f(E_{p})-f(\xi_{p-q})}{q_0+E_{p}-\xi_{p-q}}\nonumber\\
&&-\frac{1-f(E_{p})-f(\xi_{p+q})}{q_0-E_{p}-\xi_{p+q}}
+\frac{f(E_{p})-f(\xi_{p+q})}{q_0+E_{p}-\xi_{p+q}}\Bigg],
\end{eqnarray}
from which we obtain $J(q_0,{\bf 0})=0$ and
\begin{equation}
\frac{\partial^nJ(q_0,{\bf q})}{\partial q_0^n}\bigg|_{(0,{\bf
0})}=0
\end{equation}
for any integer $n$. For the derivative with respect to ${\bf q}$,
only the second derivative is nonzero,
\begin{equation}
\frac{\partial^2J(q_0,{\bf q})}{\partial {\bf q}^2}\bigg|_{(0,{\bf
0})}=-\frac{1}{m}H(0,{\bf 0}),\ \ \ \ \ \frac{\partial^nJ(q_0,{\bf
q})}{\partial {\bf q}^n}\bigg|_{(0,{\bf 0})}=0,\ \ n\neq2.
\end{equation}

\section {The functions $\Pi_I^{ij}(q)$ for $I=3$ }
\label{app2}
In this Appendix we evaluate the functions $\Pi_I^{ij}(q)$ for
$I=3$. From the relations $\Pi_3^{22}(q)=\Pi_I^{11}(-q)$ and
$\Pi_3^{21}(q)=\Pi_I^{12}(q)$, we need to evaluate $\Pi_3^{11}$
and $\Pi_3^{12}$ only. Completing the Matsubara frequency
summation and performing a shifting ${\bf p}\rightarrow {\bf
p}-{\bf q}/2$, we obtain
\begin{eqnarray}
\Pi_3^{11}(q)&=&\frac{1}{g}+\int\frac{d^3{\bf
p}}{(2\pi)^3}\Bigg[\left(\frac{u_{p-q/2}^2u_{p+q/2}^2}{q_0-E_{p-q/2}-E_{p+q/2}}
-\frac{v_{p-q/2}^2v_{p+q/2}^2}{q_0+E_{p-q/2}+E_{p+q/2}}\right)\left(1-f(E_{p-q/2})-f(E_{p+q/2})\right)\nonumber\\
&&\ \ \ \ \ \ \ \ \ \ \ \ \ \ \ \
+\left(\frac{v_{p-q/2}^2u_{p+q/2}^2}{q_0+E_{p-q/2}-E_{p+q/2}}
-\frac{u_{p-q/2}^2v_{p+q/2}^2}{q_0-E_{p-q/2}+E_{p+q/2}}\right)\left(f(E_{p-q/2})-f(E_{p+q/2})\right)\Bigg],\nonumber\\
\Pi_3^{12}(q)&=&\Delta^2\int\frac{d^3{\bf
p}}{(2\pi)^3}\Bigg[\left(\frac{1}{q_0-E_{p-q/2}-E_{p+q/2}}-\frac{1}{q_0+E_{p-q/2}+E_{p+q/2}}\right)
\frac{1-f(E_{p-q/2})-f(E_{p+q/2})}{2E_{p-q/2}E_{p+q/2}}\nonumber\\
&&\ \ \ \ \ \ \ \ \ \ \ \ \ \ \
+\left(\frac{1}{q_0+E_{p-q/2}-E_{p+q/2}}-\frac{1}{q_0-E_{p-q/2}+E_{p+q/2}}\right)\frac{f(E_{p-q/2})-f(E_{p+q/2})}{2E_{p-q/2}E_{p+q/2}}\Bigg].
\end{eqnarray}
\end{widetext}

\end{document}